%% file: conference_101719.tex
\documentclass[conference]{IEEEtran}
\IEEEoverridecommandlockouts
\usepackage{cite}
\usepackage{amsmath,amssymb,amsfonts}
\usepackage{algorithmic}
\usepackage[ruled,norelsize]{algorithm2e} % You are not supposed to use the \algorithm float
\usepackage{graphicx}
\usepackage{textcomp}
\usepackage{xcolor}
\usepackage{subcaption}
\usepackage{booktabs} % for \toprule etc.
\def\BibTeX{{\rm B\kern-.05em{\sc i\kern-.025em b}\kern-.08em
    T\kern-.1667em\lower.7ex\hbox{E}\kern-.125emX}}
\begin{document}

\title{Optimization Heuristics for Cost-Efficient Long-Term Cloud Portfolio Allocations Under Uncertainty}

\author{\IEEEauthorblockN{Maximilian Kiessler}
\IEEEauthorblockA{\textit{Faculty of Computer Science} \\
\textit{University of Vienna} \\
Vienna, Austria \\
max.kiessler@gmail.com}
\and
\IEEEauthorblockN{Valentin Haag}
\IEEEauthorblockA{\textit{Faculty of Computer Science} \\
\textit{University of Vienna}\\
Vienna, Austria \\
valentin.haag94@gmail.com}
\and
\IEEEauthorblockN{Benedikt Pittl}
\IEEEauthorblockA{\textit{Faculty of Computer Science} \\
\textit{University of Vienna}\\
Vienna, Austria \\
benedikt.pittl@univie.ac.at}
\and
\IEEEauthorblockN{Erich Schikuta}
\IEEEauthorblockA{\textit{Faculty of Computer Science} \\
\textit{University of Vienna}\\
Vienna, Austria \\
erich.schikuta@univie.ac.at}
}

%\author{\IEEEauthorblockN{Authors removed for review}
%\IEEEauthorblockA{\textit{ } \\
%  \hspace{1cm}  \\
%  \hspace{1cm} \\
%  \\
%  \\
%  \\
% }
%}

\maketitle

\begin{abstract}
Today's cloud infrastructure landscape offers a broad range of services to build and operate software applications. The myriad of options, however, has also brought along a new layer of complexity. When it comes to procuring cloud computing resources, consumers can purchase their virtual machines from different providers on different marketspaces to form so called cloud portfolios: a bundle of virtual machines whereby the virtual machines have different technical characteristics and pricing mechanisms. Thus, selecting the right server instances for a given set of applications such that the allocations are cost efficient is a non-trivial task.

In this paper we propose a formal specification of the cloud portfolio management problem that takes an application-driven approach and incorporates the nuances of the commonly encountered reserved, on-demand and spot market types. We present two distinct cost optimization heuristics for this stochastic temporal bin packing problem, one taking a naive first fit strategy, while the other is built on the concepts of genetic algorithms. The results of the evaluation show that the former optimization approach significantly outperforms the latter, both in terms of execution speeds and solution quality.
\end{abstract}


\input{text/1_introduction}
\input{text/2_state_of_the_art}
\input{text/3_problem_formulation}
\input{text/4_optimization_approaches}
\input{text/5_evaluation}
\input{text/6_conclusion}
\bibliographystyle{IEEEtran}
\bibliography{IEEEabrv,mybibliography}

\end{document}

%% file: text/1_introduction.tex
\section{Introduction}
\label{introduction}
The ever-increasing relevance of cloud computing has brought along a paradigm shift in terms of how modern software applications are developed and operated. While these advancements have certainly made reliable high-performance infrastructure more accessible, they have also added an additional layer of complexity. Heterogeneous cloud portfolio management, where the virtual machines are purchased from different marketspaces, is now concerned with the challenging task of finding cost-efficient resource allocations for a given set of applications. 

% utilizing these new services to find economical resource allocations. That is, for a given set of applications the most cost-efficient servers ought to be selected. 

Nowadays, well-known cloud service providers such as Amazon\footnote{https://aws.amazon.com/de/ec2/} commonly offer three different types of marketplaces for computational resources, i.e. the reserved, on-demand and spot marketspace. The reserved instance type offers steep discounts, but also requires longer-term commitments, as they must be purchased by the client for a predefined time period. On-demand instances, on the other hand, are characterized by their flexibility since providers usually employ a pay-per-use billing method for this type of service. This flexibility, however, incurs a higher price in comparison to the other types. Finally, on the spot marketplace one can find excess compute resources that are heavily discounted. However, instances of this type may be revoked at the service provider's discretion and are therefore only suitable for highly fault-tolerant workloads. 

%Past initiatives towards creating an aggregated cloud market such as the~\emph{Deutsche Börse Cloud Exchange (DBCE)}\footnote{http://www.picse.eu/cloud-marketplace-iaas-deutsche-börse-cloud-exchange} underpin the industry relevance of creating cost-efficient portfolios from existing Cloud marketspaces.

The heterogeneous nature of cloud markets is not the only complexity driver for this cost optimization problem. An aspect not to be neglected is the fact that applications typically are time-bound. That is, a cost-efficient resource allocation must be found over a longer temporal horizon and not only for a single point in time. Furthermore, applications rarely show a constant resource requirement. It is more common that workloads experience periods of high as well as low demands. 

To our knowledge, scientific literature does not yet provide a cloud portfolio optimization model that combines the concepts of heterogeneous marketplaces, variable resource demands, as well as the temporal character of workloads. However, all three aspects can be considered to be of integral importance to a realistic depiction of the problem, as well as any derived cost-optimization approach. Thus, with this paper we contribute to this field by 
\begin{itemize}
\item[(\emph{i})] establishing a formal problem specification that incorporates the aforementioned relevant features of cloud portfolio management, and by 
\item[(\emph{ii})] proposing and evaluating two distinct optimization heuristics for creating cost-efficient portfolio allocations.
\end{itemize}

\begin{figure}[ht]
    \centering
    \begin{subfigure}[b]{\linewidth} %nice
    \includegraphics[width=\linewidth]{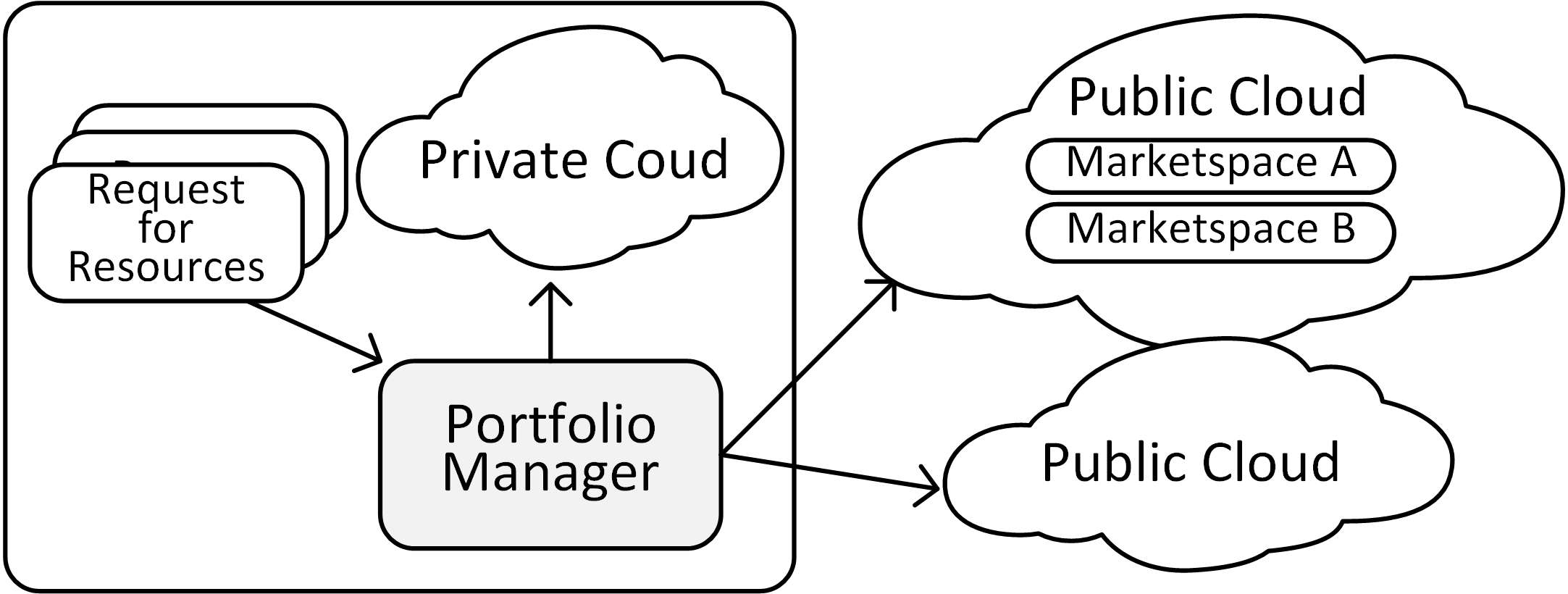}
    \end{subfigure}
\caption{\label{fig:protfoliomanager}Portfolio manager creating optimal cloud portfolios by purchasing resources from different providers and marketspaces}    
\end{figure}

The paper at hand is part of our research on an autonomous Cloud Portfolio Manager as depicted in figure~\ref{fig:protfoliomanager}. It is an entity that creates for a bundle of requested resources an optimal cloud portfolio by purchasing them from different providers as well as marketspaces. A precondition for creating such an overarching system is the determination of appropriate algorithms for creating cloud portfolios. Hence, in this paper we analyze two different cloud portfolio algorithms.

The structure of the paper is as follows: In section \ref{state_of_the_art} we give an overview of related work. Section \ref{problem_formulation} is lays out the exact mathematical model for the problem formulation. Section \ref{optimization_approaches} presents the two novel algorithmic optimization heuristics, and in section \ref{evaluation} the results of the empirical evaluation are outlined. The conclusions of the research work and future directions are discussed in section \ref{conclusion}.

%% file: text/2_state_of_the_art.tex
\section{State of the Art}
\label{state_of_the_art}
One of the key challenges of cloud portfolio management is the heterogeneous nature of the relevant marketspaces. In literature the spot market has received comparatively more research attention than the on-demand and reserved ones. This can be attributed to the fact that, depending on the respective cloud service provider, spot instances can offer significant optimization potential, due to their unique and dynamic pricing mechanisms. Sharma et al., for instance, focused on preemtible servers and utilized concepts from the domain of financial modelling to construct cost-efficient cloud portfolios while also adhering to the specific requirements of the applications that needed to be served \cite{sharma2017portfolio}. However, since spot instances are preemptible they require special mitigation procedures to avoid the loss of data upon sudden revocation. Jangjaimon and Tzeng tried to address this problem by proposing a sophisticated checkpointing mechanism \cite{jangjaimon2013effective}. The process of checkpointing, i.e. regularly persisting a recoverable state for computational workloads, is quite common when working with transient cloud servers, as also pointed out by Tang et al. In their work the authors propose a biding strategy for spot instances based on the concepts of a constrained Markov decision process. Their research focused on the EC2 Spot market of Amazon Web Services (AWS) and aimed to provide a balanced approach between cost-optimization and revocation mitigation \cite{tang2012towards}. However, as mentioned by Baughman et al., AWS revised the pricing mechanism of its preemptible servers in 2017. The currently active schema makes prices less volatile, meaning that instead of having high price peaks according to the fluctuating supply and demand, the curve is now flattened \cite{baughman2019deconstructing}. Based on this overhauled pricing mechanism of EC2 spot instances, Zhou et al. developed a long-term price prediction and dynamic task reallocation system. However, their scheduling system again only focuses on the spot market and does not take into account a heterogeneous cloud portfolio \cite{zhou2021farspot}.

Mireslami et al., on the other hand, incorporated on-demand and reserved instances in their provisioning approach. However, besides the fact that the authors did not consider the spot market, their model also lacks a sophisticated temporal component and thus is not suitable for long-term task scheduling \cite{mireslami2019dynamic}. Li et al. do include time-bound constraints in their dynamic bin packing approach, however, while they do assume that the size of an item may vary from one to another, they do not model the capacity requirements as a random variable. Furthermore, they also do not fully incorporate the heterogeneous nature of cloud markets \cite{li2015dynamic}. Nodari et al. also do not pay attention to the spot market, but instead put an emphasis on reserved and on-demand instances in their stochastic model, to which they apply the concepts of inventory theory \cite{nodari2016inventory}. In the research of Haussmann et al. a cost model was presented that incorporated volatile as well as secure instance types. Their research showed that preemptible spot instances can contribute to significantly lower cost in the domain of high performance parallel computing \cite{haussmann2019cost}. However, while their model does consider varying execution times, it too is not suitable for long-term job scheduling. Alenizi et al. try to capture the time-bounded nature of workflows by making the remaining expected execution time an integral part of their model. Their approach, however, again only considers reserved and on-demand instances. Further, it also does not consider various starting times of workflows and thus is more geared towards dynamic ad hoc resource allocations \cite{alenizi2020cost}. Shen et al. do consider separate arrival and departure times for workflows in their job scheduling model, but also do not incorporate all three of the aforementioned cloud markets. Further, their integer programming problem formulation only considers non-fluctuating resource demands for the individual workflows \cite{shen2013scheduling}. A more holistic approach to cloud portfolio management was taken by Pittl et al., who showed that heterogeneous portfolios tend to be more cost-efficient. Furthermore, they also pointed out the importance of right-sizing the server instances to the capacity requirements of the workloads \cite{pittl2019cost}.

The task of right-sizing is also highly relevant to the closely-related domain of energy-efficient data center operations. For this field Hwang and Pedram proposed a portfolio-based optimization approach with a probabilistic model, where the assigned workloads have resource demands that are characterised by a normal distribution \cite{hwang2012portfolio}. Martinovic et al. also use a problem formulation based on probabilities rather than deterministic aspects. They then applied a stochastic bin packing approach to find efficient server allocations \cite{martinovic2020stochastic}. Fatima et al. evaluated a particle swarm optimization algorithm against other approaches in a scenario where the server capacities are assumed to be variable \cite{fatima2018virtual}. Wu et al., on the other hand, used a more traditional genetic algorithm concept in their proposed solution for physical and virtual server instance consolidation \cite{wu2012energy}. De Cauwer et al. assumed deterministic resource demands for their tasks, which arguably is a less plausible depiction of reality. However, they also introduced a temporal component to their problem formulation. Instead of simply finding efficient server consolidation schemes for a single point in time, they added time as another resource dimensions. This better captures the true nature of many computational workloads, which often do not run indefinitely, but rather have a predefined lifespan \cite{de2016temporal}.

As has been shown, while one can find extensive research on the subject of cloud cost optimization, very little work has been done on long-term allocations. Long-term cost-efficient cloud portfolio management does require to put an emphasis on the temporal component that is inherent to such a problem. Some research tries to capture this time-bound characteristic \cite{li2015dynamic, haussmann2019cost, alenizi2020cost, shen2013scheduling, de2016temporal}, however, to the best of our knowledge their scientific literature does not yet provide optimization approaches that adequately capture this constraint, while also incorporating the heterogeneous nature of cloud markets. As mentioned, the research by Pittl et al. performs a more holistic analysis of resource management, i.e. procurement options from reserved, on-demand as well as spot markets are considered, but it can be argued that it does not appropriately address the uncertain nature of capacity demands \cite{pittl2019cost}. As pointed out, researcher that already previously incorporated stochastic problem formulations in their models \cite{nodari2016inventory, martinovic2020stochastic}, is not suited for the task of long-term resource allocations in heterogeneous cloud market environments.

Based on our past research on resource optimization techniques \cite{schikuta08wfopt,pittl15resalloc} and cloud market analysis \cite{pittl16simresalloc,mach12contracting} we try to close this gap. That is, our model not only considers reserved, on-demand and spot instances, it also puts an emphasis on long-term resource allocations. The proposed model and the derived optimization approaches support fine-grained task scheduling over a longer planning horizon, which allows to properly make use of reserved instances. Furthermore, our application driven approach is based on the assumption of uncertain resource requirements and thus also adds a stochastic notion to the problem formulation. 

%% file: text/3_problem_formulation.tex
\section{Problem Formulation}
\label{problem_formulation}

The problem formulation for cost-efficient cloud portfolio management presented in this section uses the notation as defined in table \ref{tab:problem_notation}.

\begin{table}[ht]
\renewcommand{\arraystretch}{1.3}
\begin{center}
\caption{\label{tab:problem_notation}Notation for Problem Formulation}

\begin{tabular}{cl}
 %\hline
 \toprule
 \multicolumn{1}{p{1.5cm}}{\centering \textbf{Parameter}} & 
 \multicolumn{1}{p{6cm}}{\centering \textbf{Description}} \\
 \midrule
 $I$ & A set of selected cloud instances (hosts) \\ 

 $A$ & A set of applications \\

 $T$ & A set of time slots for which to optimize \\
 
 $x_{ait}$ & Variable denoting assignments of apps to instances \\
 
 $S_a$ & The starting time of app $a$  \\
 
 $F_a$ & The finishing time of app $a$ \\
 
 $U_a$ & Indicates if app $a$ is preemptible \\

 $\mu_a$ & The expected resource demand of app $a$ \\
 
 $\sigma_a$ & The std. deviation of the resource demand of app $a$ \\

 $R_i$ & The resource capacity of instance $i$  \\

 $C_i$ & The cost of instance $i$ for one time slot \\

 $B_i$ & The first available time slot of an instance \\

 $E_i$ & The last available time slot of an instance \\
 
 $O_i$ & Marks instances as suitable for non-preemptible apps \\

 $D_{it}$ & Aggregated resource demand of instance $i$ at time $t$ \\
 
 $Q_{min}$ & The desired quality of service \\
 
 \bottomrule
\end{tabular}

\end{center}
\end{table}

In the scope of this work a cloud portfolio is defined as a set of \textit{cloud instances} $I$ that are necessary to host a set of \textit{applications} $A$. The optimization problem, in short, is then concerned with finding a cost-efficient allocation of these applications and instances. 

Applications running in the cloud are typically characterized by their resource demands. Following the proposed approach of Hwang and Pedram these capacity requirements are assumed to be fluctuating and are therefore represented by an expected demand mean $\mu_{a}$ and corresponding standard deviation $\sigma_{a}$ \cite{hwang2012portfolio}. Furthermore, it is assumed that workloads have  varying life spans, which is denoted by the starting time $S_a$ and finishing time $F_a$ of an application. Note that $S_a < F_a$ must apply. Since spot instances are only suitable for very fault-tolerant processes, the binary variable $U_a$ is used to denote whether or not an application can be assigned to such a preemptible server instance.

\begin{align}
U_a = \begin{cases}
1 & \text{if $a$ is preemptible} \\
0 & \text{else}
\end{cases}
\label{eqn:preemptible_applications}
\end{align}

An instance $i \in I$ has a predefined resource capacity $R_i$ (e.g. CPU, RAM). Moreover, each type of instance is also associated with a certain price per time unit $C_i$. The total cost incurred by a single host is based on the price per time slot and the overall up-time. The up-time of an instance is given by its starting time $B_i$ and ending time $E_i$, where $B_i < E_i$ again must hold true. To denote if an instance is considered to be a spot (preemptible) type, the binary parameter $O_i$ is used.

\begin{align}
O_i = \begin{cases}
1 & \text{if $i$ is only suitable for preemptible apps} \\
0 & \text{else}
\end{cases}
\label{eqn:preemptible_instances}
\end{align}

The aggregated demand of all applications assigned to instance $i \in I$ at time slot $t \in T$ is represented by the parameter $D_{it}$. This aggregated demand is also a random variable and thus one can only evaluate the probability with which an instance stays within the designated capacity limits. The proposed model assumes that one can specify a desired minimum quality of service, denoted by $Q_{min}$. In case the probability that the aggregated demand $D_{it}$ stays below the provided capacity of the instance $R_i$ does not satisfy the minimum quality of service $Q_{min}$ for time slot $t$, then the allocation is invalid. 

To formally model the application and instance assignments, while also considering the temporal restrictions of the problem, the approach suggested by Dell'Amico et al. is used. The variable $x$ denotes which hosting instance an application has been assigned to at a specific point in time. The resource demands of an application $a \in A$ are considered to be $0$ for any time slot $t \in T$ if $t < S_a$ or $t > F_a$ \cite{dell2020branch}.

\begin{align}
x_{ait} = \begin{cases}
1 & \text{if $a$ is assigned to $i$ at time slot $t$} \\
0 & \text{else}
\end{cases}
\label{eqn:assignment_variable}
\end{align}

The cloud portfolio management problem can now be defined as follows:

\begin{equation}
    \label{eqn:min_portfolio_cost}
    min \sum_{i \in I} C_i * (E_i - B_i)
\end{equation}

\begin{align}
          \label{eqn:assign_app_to_instance}
    s. t. &\sum_{i \in I} x_{ait} = 1           &   &   &   &\forall a \in A, \forall t \in [S_{a}, F_{a}]\\
          \label{eqn:suitable_host}
          &\sum_{i \in I} x_{ait} * U_a \geq O_i  &   &   &   &\forall a \in A, \forall t \in [S_{a}, F_{a}] \\
          \label{eqn:resource_constraint}
          &P(D_{it} < R_i) \geq Q_{min}            &   &   &   &\forall i \in I, \forall t \in [B_{i}, E_{i}]\\
          \label{eqn:assignment_range}
          &x_{ait} \in \{0, 1\}                 &   &   &   & \\
          \label{eqn:preemptible_application_range}
          &U_a \in \{0, 1\}                     &   &   &   & \\
          \label{eqn:preemptible_instance_range}
          &O_i \in \{0, 1\}                     &   &   &   & \\
          \label{eqn:quality_range}
          &Q_{min} \in [0, 1]                     &   &   &   &
\end{align}

Constraint \ref{eqn:assign_app_to_instance} states that the optimization approach needs to assign each application to an instance. At any given point in the up-time of an application, only one instance can be the dedicated host. Constraint \ref{eqn:suitable_host} requires the selected host to be from a suitable marketspace. Furthermore, equation \ref{eqn:resource_constraint} stipulates that for each time slot an instance is active, the probability that the resource demand stays within the capacity limits of the respective instance is at least the predefined quality of service. This constraint is based on the approach proposed by Hwang and Pedram \cite{hwang2012portfolio} and assumes applications with load profiles that do not show any correlation and have a normal distribution. With constraints \ref{eqn:assignment_range} to \ref{eqn:quality_range} the auxiliary variables are bound to a valid range. 

%% file: text/4_optimization_approaches.tex
\section{Optimization Approaches}
\label{optimization_approaches}
The cloud portfolio optimization problem, as modeled in section \ref{problem_formulation}, is in essence a multi-dimensional packing problem. Such problems are NP-hard and thus complex to solve \cite{chekuri2004multidimensional}. While exact algorithms for much simpler variations of the bin packing problem have been proposed in literature - for instance by Martello and Vigo \cite{martello1998exact} - a complete enumeration of all possible solutions is often not computationally feasible. Hence, in this work we present two distinct optimization heuristics, which try to find good approximations of optimal solutions. 

\subsection{The Greedy Optimization Approach}
\label{greedy_approach}

The first of the two presented optimization heuristics incorporates principles from the well-known first fit decreasing (FFD) approach \cite{man1996approximation}, as well as the proposed portfolio management strategy by Hwang and Pedram \cite{hwang2012portfolio}. The algorithm outlined in the following, which has been named \emph{Efficient Resource Inference for Cloud Hosting} (ERICH), consists of 4 stages:

\textbf{Stage 1}: The algorithm expects sets of preemptible and non-preemptible applications, as well as all possible instance types from the three respective marketspaces as input. In the initialization phase applications are sorted according to increasing starting dates and non-increasing resource demand standard deviations. This is based on the proposed idea by Hwang and Pedram, who suggested that grouping together workloads with similar resource demand deviation may lead to overall reduced capacity needs \cite{hwang2012portfolio}. Furthermore, the various instance types are sorted by cost per time slot for the provided capacity in an increasing order, ensuring that the algorithm chooses cost-efficient hosts first.

\textbf{Stage 2}: According to the first fit decreasing approach, the algorithm tries to fit all non-preemptible applications into reserved instances. For each application, it is first checked if a suitable host, i.e. one covering the entire life cycle of the application while also providing the required capacity, already exists in the constructed portfolio. If it does, then the application is assigned to said instance. In case no candidate host exists, a new one is allocated based off the first instance type satisfying the application's requirements. 

\textbf{Stage 3}: Although reserved instances usually provide steep discounts compared to the on-demand marketspace, the fact that they have a minimum allocation period means that \emph{stage 2} may result in a cost-inefficient portfolio. To achieve a potentially higher packing density the algorithm iterates over each allocated reserved instance. During each iteration, a new temporary portfolio without the respective reserved instance is created. Applications from the removed reserved host are then assigned to on-demand instances following the same first fit decreasing approach as described in \emph{stage 2}. If the newly created portfolio is more cost efficient, than it replaces the old one in the next iteration.

\textbf{Stage 4}: In the final step, the algorithm is concerned with finding well-fitting allocations for all preemptible applications. Preemptible workloads may be assigned to multiple hosts during their life cycle, which is why the optimization approach allocates these applications not as a whole, but rather on a individual time slot basis. That is, the algorithm first finds all time steps for which a suitable candidate hosts exists and assigns the app to these instances for the respective periods. In case there are still some assignment gaps in the schedule of the workload the algorithm proceeds with allocating new instances from the input set of spot hosts. This approach ensures that preemptible applications fill up any left-over capacity before new bins are opened. When following these scheme one may need to consider if the applications require a checkpointing mechanism similar to what has been discussed in related literature \cite{jangjaimon2013effective, tang2012towards}.

Algorithm \ref{alg:efficient_resource_inference_for_cloud_hosting} outlines the above described approach in a condensed form. The evaluation whether or not an application fits into an instance is performed based on equation \ref{eqn:resource_constraint}.

\makeatletter
\newcommand{\removelatexerror}{\let\@latex@error\@gobble}
\makeatother

\begin{figure}[!t]
\removelatexerror
\begin{algorithm}[H]
\caption{Efficient Resource Inference for Cloud Hosting}

\SetAlgoLined
\SetKwInOut{Input}{Input}
\Input{A set of non-preemptible apps $A1$; A set of preemptible apps $A2$; A set of reserved instance types $RES$; A set of on-demand instance types $ON$; A set of spot instance types $SPOT$}
\KwResult{Packing pattern $portfolio$}

sort applications $A1$ and $A2$ by increasing start time and non-increasing $\sigma{_a}$\\
sort $RES$, $ON$ and $SPOT$ by increasing $C_i$ per $R_i$ and time slot\\
$portfolio$ $\gets$ empty allocation variable\\

\ForAll{$a \in A1$}{
    assign $a$ to $portfolio$ (FFD) while only considering $RES$ instances\\
} 

\ForAll{$i \in $ reserved instances from $portfolio$}{
    $tmp\_portfolio$ $\gets$ copy of $portfolio$ without instance $i$\\
    \ForAll{$a \in i$} {
        reinsert $a$ into $tmp\_portfolio$ (FFD) including $ON$ instances\\
    }
    \If{total cost of $tmp\_portfolio <$ total cost of $portfolio$}{
        $portfolio \gets tmp\_portfolio$\\
    }
}

\ForAll{$a \in A2$}{
    assign $a$ to $portfolio$ without allocating new instances\\
    $gaps \gets$ consecutive time slots where $a$ is not yet assigned to $portfolio$\\
    \ForAll{$gap \in gaps$}{
        assign $a$ to $portfolio$ for time slots in $gap$ by allocating $SPOT$ hosts\\
    }
}

\end{algorithm}
\caption{Greedy Algorithm ERICH}
\label{alg:efficient_resource_inference_for_cloud_hosting}
\end{figure}

\subsection{The Evolutionary Optimization Approach}
\label{genetic_approach}

The application of genetic algorithms (GA) to bin packing scenarios is not a new notion. In fact, GAs have proven to be well suited to tackle the complex nature of such combinatorial optimization problems \cite{reeves1996hybrid, falkenauer1996hybrid, quiroz2015grouping, kang2012hybrid}. Their flexible framework built around genetic operators that can be adapted to the problem at hand means that GAs allow for a very efficient and targeted search in the problem space. The algorithm presented here has been named \emph{Genetic Optimization of Resource Groupings} (GEORG). In the following its most essential building blocks and genetic operators are presented.

\begin{figure}[!t]
\removelatexerror
\begin{algorithm}[H]
\caption{Genetic Optimization of Resource Groupings}

\SetAlgoLined
\SetKwInOut{Input}{Input}
\Input{A set of non-preemptible apps $A1$; A set of preemptible apps $A2$; A set of reserved instance types $RES$; A set of on-demand instance types $ON$; A set of spot instance types $SPOT$}
\KwResult{List of packing patterns (portfolios) $population$}

$population$ $\gets$ use semi-random heuristic to create initial population\\
\While{termination criteria are not met}{
    $parents$ $\gets$ fitness-based selection of individuals from $population$\\
    $offspring$ $\gets$ apply temporal biased crossover for each tuple in $parents$\\
    $offspring$ $\gets$ repair broken chromosomes in $offspring$ after crossover\\
    $offspring$ $\gets$ apply domination mutation operator to random $offspring$\\
    $offspring$ $\gets$ repair broken chromosomes in $offspring$ after mutation\\
    $population$ $\gets$ fitness-based merge of $offspring$ and current $population$\\
}
\end{algorithm}
\caption{Genetic Algorithm GEORG}
\label{alg:genetic_optimization_of_resource_groupings}
\end{figure}

\textbf{Encoding Scheme}: For bin packing problems the grouping of items and their respective bins is an essential piece of information, which is also relevant to the individual genetic operators of a GA \cite{falkenauer1996hybrid}. Falkenauer has proposed an encoding scheme, in which the chromosome is an array of bins, each holding a set of items. However, this approach is not well suited for the temporal component of the problem specification. Thus, we propose a temporal group encoding, in which each locus of the chromosome corresponds to a specific time step, with the allele being a set of instances running at the respective point in time. Every host is then mapped to a set of applications, which are assigned to the respective instance at this time slot.

\textbf{Population Initialization}: To ensure a high degree of genetic diversity while also providing initial individuals with comparatively high fitness, a hybrid approach is taken for the creation of an initial set of solutions. Fifty percent of the time applications are assigned in a manner that reduces the number of allocated instances and therefore the overall cost. In the remaining cases, the assignment is at random.

\textbf{Fitness Evaluation}: To evaluate the fitness of an individual, i.e. the quality of a solution, equation \ref{eqn:min_portfolio_cost} is used.

\textbf{Crossover}: In each generation a group of individuals is selected based on the fitness proportionate roulette wheel method. The crossover is then applied to a set of two individuals (parents) in order to yield a new offspring. We propose a biased temporal crossover operator that is built on the concepts presented by Quiroz-Castellanos et al. \cite{quiroz2015grouping} and promotes the transmission of well-fitting genetic material to the next generation. For each time step (gene), active instances are sorted in decreasing order based on their average rate of capacity utilization and cost per time slot. A partial solution is created by zip-merging the instances of both parents. The ranked partial solution is pruned by eliminating hosts of already assigned applications. Instances that result in packing patterns violating the constraints of the problem formulation are ignored in the crossover process of further time steps. The culling of instances may break the chromosome, since applications may end up being partially or fully without any host assignments. To repair the chromosome, a simple heuristic is applied to reinsert any applications that are now considered to be free.

\textbf{Mutation}: The mutation operator is applied to newly created offspring at random. Its goal is to introduce new genetic characteristics to individuals of the population, while also following a targeted approach to enhance the overall fitness. Martello and Toth \cite{martello1990lower} introduced the concept of dominance, which can be used to replace a subset of items with an item of larger or equal size. This approach has already been employed in the context of genetic algorithms by other researchers \cite{falkenauer1996hybrid, quiroz2015grouping}. As Quiroz-Castellanos et al. also pointed out, the flexibility gained by only having to reinsert smaller items can ultimately lead to a tighter packing pattern and an overall better solution. Since random chance determines which items are initially chosen to perform a dominance check, the design allows for the introduction of new variations of genetic material, while also potentially improving the already existing solution \cite{quiroz2015grouping}.

\begin{figure}[ht]
    \centering
    \begin{subfigure}[b]{\linewidth}
    \includegraphics[width=\linewidth]{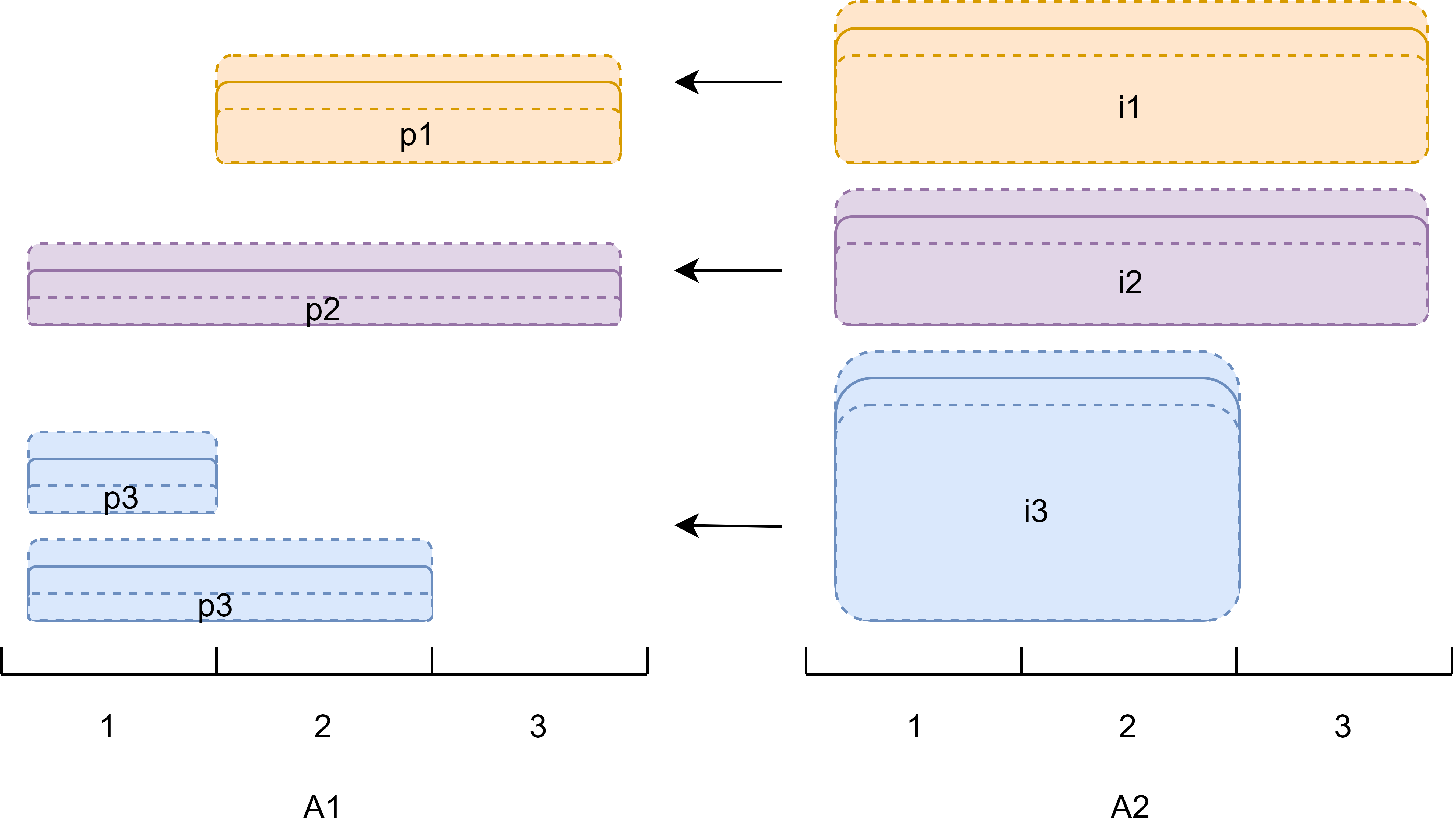}
    \end{subfigure}
\caption{\label{fig:dominance_criterion_fulfilled}Fulfilled Temporal Dominance Criterion}
\caption*{Note. Figure inspired by Falkenauer.}
\end{figure}

\begin{figure}[ht]
    \centering
    \begin{subfigure}[b]{\linewidth}
    \includegraphics[width=\linewidth]{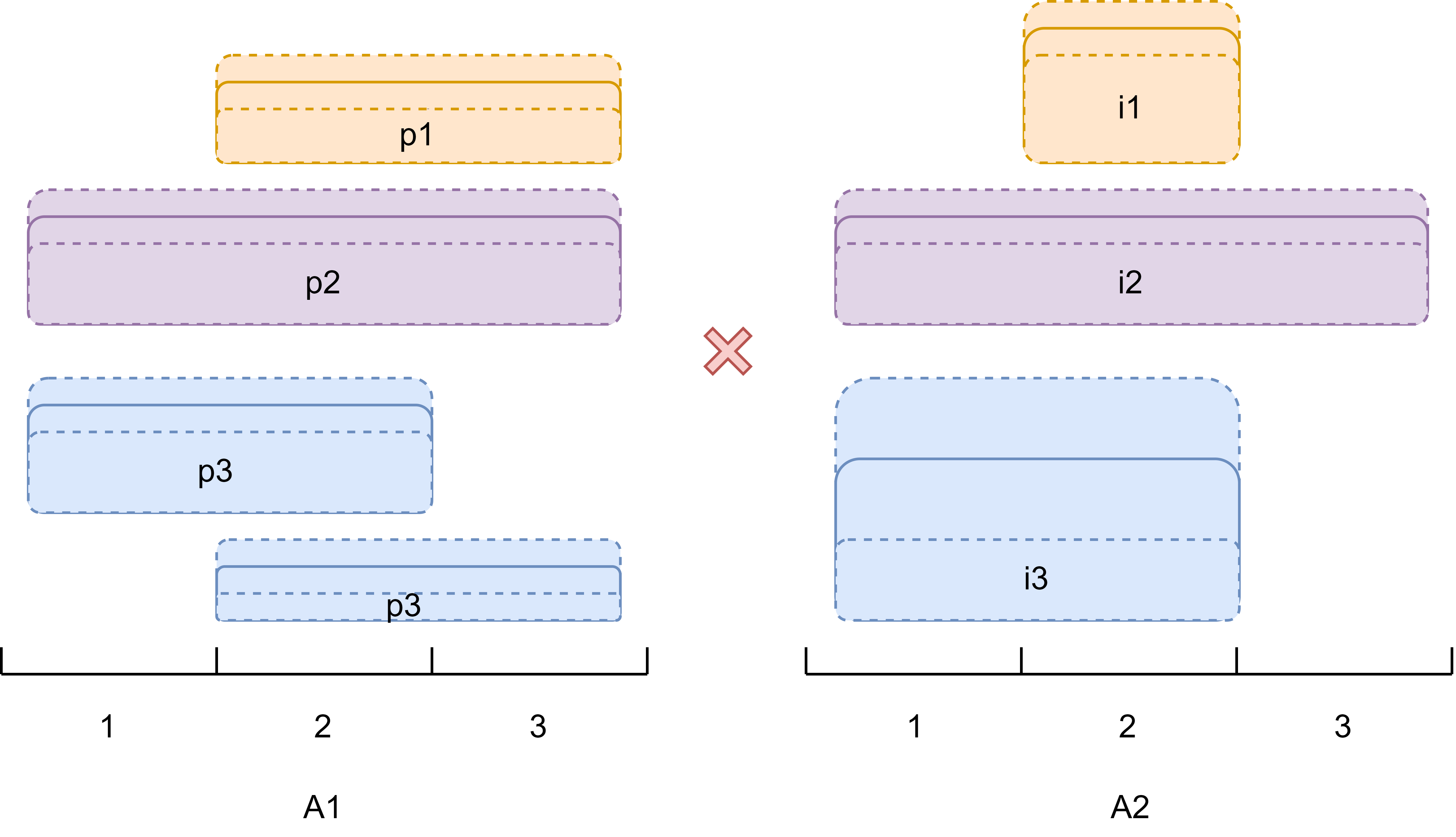}
    \end{subfigure}
\caption{\label{fig:dominance_criterion_unfulfilled}Unfulfilled Temporal Dominance Criterion}
\caption*{Note. Figure inspired by Falkenauer.}
\end{figure}

The definition of the dominance criterion needs to be slightly adapted to be applied to the cloud portfolio optimization problem with its temporal component. Furthermore, instead of comparing entire allocations, i.e. instances, with each other, the dominance definition used in the following focuses on the application level. An application $a$ that is intended to be hosted by instance $i \in I$ for the time slots $[S_d, F_d]$ is said to dominate a partition of apps $P$ from instance $i$ if the time span denoted by $[S_d, F_d]$ encompasses all assignments slots of the partition for the respective host. Furthermore, the expected resource demands of the dominating application must be greater than the aggregated capacity requirements of all elements of the dominated partition. In fact, the probability that the resource demands of the dominating app exceed those of the dominated workloads must be greater than fifty percent. The depictions in figure \ref{fig:dominance_criterion_fulfilled} and \ref{fig:dominance_criterion_unfulfilled} show the application level resource requirements over a horizon of three time steps. The bold lines indicate the expected resource demands, while the dotted ones showcase the deviations. 

Figure \ref{fig:dominance_criterion_fulfilled} depicts a set of three applications and a set of three partitions. Each individual application in $A2$ dominates the respective partition of $A1$ in the same color. The above listed criteria for dominance are fulfilled in all dimensions, i.e. capacity and time. For instance, application $i1$ clearly has a higher expected resource demand than partition $p1$. At the same time, $i1$ also dominates $p1$ in the temporal dimension. The application $i2$ matches the starting and finishing times of partition $p2$ exactly. This is in accordance with the temporal dominance criterion. It further also has a higher expected resource demand and thus it can be said that $i2$ dominates $p2$. The partition $p3$ consists of two applications, with a combined expected resource demand that is less than that of $i3$. Since $i3$ also covers at least the same time slots as $p3$ it can be said that $i3$ dominates $p3$. The mutation operator now could replace one of the dominated partitions and yield a tighter packing pattern. Figure \ref{fig:dominance_criterion_unfulfilled} shows a scenario, in which non of the selected applications dominates any partition. For instance, while $i1$ certainly has a higher expected resource demand, it does not cover the entire live span of partition $p1$. On the other hand $i2$ may cover the same time period as $p2$, but has the same expected resource demands. Hence, it too cannot be considered as a dominating application. In the case of application $i3$, the deviations from the expected capacity requirements are significant. Therefore, the probability that $i3$ experiences higher resource demands than what is the expected aggregated distribution for partition $p3$ is quite low. Besides, the temporal constraints is not satisfied either, which means that $i3$ cannot be considered to dominate $p3$.

The temporal dominance criterion is checked against partitions of applications of size two for the respective candidate instance, as otherwise the computational efforts might become unfeasible \cite{quiroz2015grouping}. In case the dominance criterion holds true, the unassigned application is swapped with the partition. Similar to the crossover operator, this procedure may leave certain applications unassigned, meaning that the chromosome no longer represents a valid solution. In such a case the same aforementioned insertion heuristic is now again used to repair the corrupted individual, as otherwise the genetic algorithm would not be able to continue its process.

\textbf{Insertion Heuristic}: Equation \ref{eqn:assign_app_to_instance} of the problem specification essentially requires that each application must be assigned an appropriate host for all relevant time slots. During the crossover and mutation processes this constraint might end up being violated, resulting in a broken chromosome. Preemptible applications might have multiple gaps in their packing pattern, where no valid assignment is available. In any case, the insertion heuristic follows a naive first-fit approach, where for each assignment gap of an orphaned application a candidate instance from the existing chromosome is selected. If no candidate host exists, then a new random instance is created to host the respective application. The component of randomness to this rather simple heuristic is intended to mitigate the risk of premature converges on local optima by ensuring that the genetic diversity is kept at an adequate level.

\textbf{Termination}: Commonly encountered stopping criteria can be employed, such as a maximum number of processed generations or a certain threshold when it comes to the convergence level of the fitness scores among individuals in the population \cite{safe2004stopping}.

%% file: text/5_evaluation.tex
\section{Evaluation}
\label{evaluation}
The proposed optimization heuristics have been empirically evaluated with synthetic data. The algorithms were implemented using Python 3.9 and the tests were run on a Windows machine with an Intel Core i7-4770 processor (3.4 GHz base clock, 3.9 GHz turbo) and 16 GB of DDR3 memory at 1600 MHz.

\subsection{Data Set Description}
\label{data_set_description}
With the problem at hand, there are multiple dimensions which contribute to the difficulty of a particular test set. As is usually the case for bin packing optimization algorithms, the greater the number of items to be assigned the more challenging a task becomes. However, due to the temporal component of the problem formulation the number of allocation periods for which to find suitable assignments must not be neglected. It was therefore thought to be imperative that the variety in the data reflects these difficulty attributes.

\begin{table}[ht]
\renewcommand{\arraystretch}{1.3}
\caption{\label{tab:application_data}Summary of Application Data Sets}
\setlength{\tabcolsep}{5pt}
\begin{center}
\begin{tabular}{lrrrrrrrrr}

\toprule
\multicolumn{1}{p{0.2cm}}{\centering \textbf{App. \\ Set}} &
\multicolumn{1}{p{0.2cm}}{\centering \textbf{Non-Pre.}} &
\multicolumn{1}{p{0.2cm}}{\centering \textbf{Pre.}} &
\multicolumn{1}{p{0.2cm}}{\centering \textbf{Avg. \\ Res. \\ Dem.}} &
\multicolumn{1}{p{0.2cm}}{\centering \textbf{Std. \\ Res. \\ Dem.}} &
\multicolumn{1}{p{0.2cm}}{\centering \textbf{Avg. \\ Res. \\ Dev.}} &
\multicolumn{1}{p{0.2cm}}{\centering \textbf{Std. \\ Res. \\ Dev.}} &
\multicolumn{1}{p{0.4cm}}{\centering \textbf{Avg. \\ Alloc. \\ Periods}} &
\multicolumn{1}{p{0.4cm}}{\centering \textbf{Std. \\ Alloc. \\ Periods}} &
\\
\midrule
apps\_1 & 14 & 6 & 3.27 & 1.71 & 0.53 & 0.48 & 43.15 & 33.4 \\
apps\_2 & 59 & 41 & 3.0 & 2.62 & 0.53 & 0.74 & 63.93 & 43.94 \\
apps\_3 & 10 & 10 & 3.02 & 2.01 & 0.71 & 0.63 & 212.2 & 167.88 \\
apps\_4 & 42 & 58 & 3.1 & 2.57 & 0.5 & 0.59 & 237.16 & 171.57 \\
apps\_5 & 7 & 13 & 3.12 & 2.69 & 0.6 & 0.56 & 2758.55 & 1996.98 \\
apps\_6 & 41 & 59 & 2.78 & 1.97 & 0.49 & 0.57 & 2871.74 & 2055.67 \\

\bottomrule

\end{tabular}
\end{center}

\end{table}

While the data was synthetically created, careful attention was paid to crafting realistic scenarios. That is, the expected price discounts for spot and reserved instances relative to on-demand ones are based on real-world observations. %maybe mention where those observations have been made
Furthermore, the price to capacity relations are also modeled according to commonly encountered offerings of popular cloud service providers. Table \ref{tab:application_data} summarizes the key resource demand and allocation characteristics of the application data sets. Note that the columns \emph{Non-pre.} and \emph{Pre.} list the number of non-preemptible and preemptible applications that each data set holds. The table further describes the expected average resource demands (\emph{Avg. Res. Dem.}) as well as the deviations from this average within the data set (\emph{Std. Res. Dem.}). Moreover, the column \emph{Avg. Res. Dev.} highlights the mean resource demand deviations that are assumed for applications. With column \emph{Std. Res. Dev.} the fluctuations of these deviations within the data set are described. Furthermore, table \ref{tab:application_data} also lists the average allocation periods.

These samples were combined with the instance type data sets described in table \ref{tab:instance_type_data}. Each instance type data set included five hundred different types from which the algorithms could choose. Table \ref{tab:instance_type_data} describes the mean capacities of the instance types, as well as the average prices per market space. It is important to note that the high deviations of the cost levels highlight the great price diversity found in the data sets. The following six unique test cases were created: $case\_1$ ($apps\_1$, $types\_1$), $case\_2$ ($apps\_2$, $types\_1$), $case\_3$ ($apps\_3$, $types\_2$), $case\_4$ ($apps\_4$, $types\_2$), $case\_5$ ($apps\_5$, $types\_3$) and $case\_6$ ($apps\_6$, $types\_3$). The full data sets are available publicly\footnote{https://gitlab.com/MFJK/optimization-heuristics-for-cost-efficient-cloud-resource-allocations}.

\begin{table}[ht]
\setlength{\tabcolsep}{5pt}
\caption{\label{tab:instance_type_data}Summary of Instance Type Data Sets}
\begin{center}
\renewcommand{\arraystretch}{1.3}
\begin{tabular}{lrrrrrrrrrr}
\toprule
\multicolumn{1}{p{0.2cm}}{\centering \textbf{Instance \\ Type \\ Set}} &
\multicolumn{1}{p{0.2cm}}{\centering \textbf{Avg. \\ Cap.}} &
\multicolumn{1}{p{0.2cm}}{\centering \textbf{Std. \\ Cap.}} &
\multicolumn{1}{p{0.2cm}}{\centering \textbf{Avg. \\ Res. \\ Prc.}} &
\multicolumn{1}{p{0.2cm}}{\centering \textbf{Std. \\ Res. \\ Prc.}} &
\multicolumn{1}{p{0.2cm}}{\centering \textbf{Avg. \\ On. \\ Prc.}} &
\multicolumn{1}{p{0.2cm}}{\centering \textbf{Std. \\ On. \\ Prc.}} &
\multicolumn{1}{p{0.2cm}}{\centering \textbf{Avg. \\ Spot \\ Prc.}} &
\multicolumn{1}{p{0.2cm}}{\centering \textbf{Std. \\ Spot \\ Prc.}} &

\\
\midrule
types\_1 & 9.60 & 8.77 & 2.27 & 2.85 & 3.10 & 2.16 & 2.53 & 2.21 \\
types\_2 & 10.32 & 11.40 & 2.16 & 2.38 & 3.13 & 2.57 & 3.15 & 4.84 \\
types\_3 & 9.79 & 9.91 & 2.41 & 3.80 & 3.10 & 2.41 & 2.33 & 1.74 \\

\bottomrule
\end{tabular}
\end{center}

\end{table}

\subsection{Results}
\label{results}
The evaluation was conducted with three quality criteria in mind, i.e. execution speed, packing density and overall cost of the constructed cloud portfolios. In the following the results are discussed.

\begin{figure}[ht]
    \centering
    \begin{subfigure}[b]{\linewidth}
    \includegraphics[width=\linewidth]{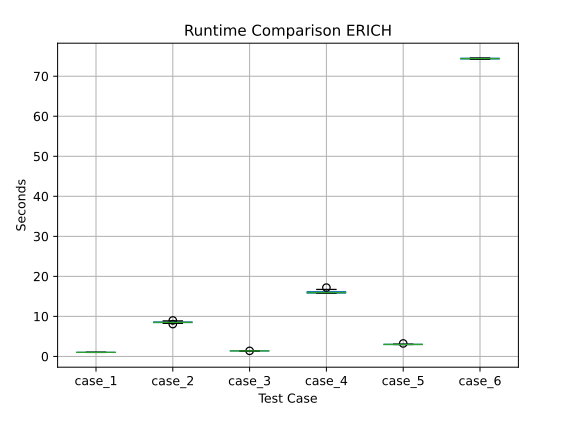}
    \end{subfigure}
\caption{\label{fig:erich_execution_speeds}Execution Speed ERICH}
\end{figure}

Since wall clock time was chosen as the evaluation metric the test cases for the evaluation of the execution speeds were run a total of 10 times to eliminate any potential side effects of background tasks. As can be seen in figure \ref{fig:erich_execution_speeds}, the execution speeds for the greedy optimization approach \emph{ERICH} was significantly better across the board. Its deterministic character resulted in more or less static runtime performance. The little variations in the execution speeds can be explained by the expected noise caused by various background tasks on the host machine.

% \begin{figure}[ht]
%     \centering
%     \begin{subfigure}[b]{.49\linewidth}
%     \includesvg[width=\linewidth]{images/erich_run_time.svg}
%     \caption{Execution speed ERICH}
%     \label{fig:erich_execution_speeds}
%     \end{subfigure}
%     \begin{subfigure}[b]{.49\linewidth}
%     \includesvg[width=\linewidth]{images/georg_run_time.svg}
%     \caption{Execution speed GEORG}
%     \label{fig:georg_execution_speeds}
%     \end{subfigure}

% \caption{\label{fig:execution_speeds}Performance in terms of execution speeds}
% \end{figure}

\begin{figure}[ht]
    \centering
    \begin{subfigure}[b]{\linewidth}
    \includegraphics[width=\linewidth]{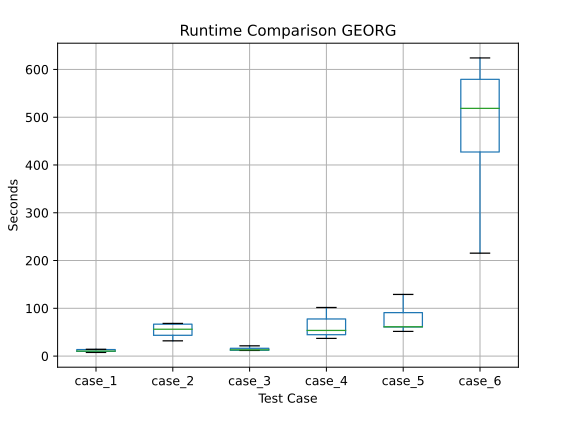}
    \end{subfigure}
\caption{\label{fig:georg_execution_speeds}Execution Speed GEORG}
\end{figure}

However, the execution speed of the GA, which can be seen in figure \ref{fig:georg_execution_speeds}, has proven to be highly volatile, which can be attributed to the fact that the individual genetic operators are heavily influenced by randomness. Furthermore, the overall runtime for each individual test case is significantly higher than the corresponding evaluation result for \emph{ERICH}. The processing time required by \emph{GEORG} is at times more than five-fold higher. The most significant factor that contributes to this is likely the crossover operator described. The temporal component of the problem specification requires a substantial amount of evaluations to be performed to yield a single offspring. Furthermore, it can easily be the case that the crossover results in certain applications not being assigned to any hosts, in which case the insertion heuristic has to reinsert these applications. The same reasoning can be applied to the mutation operator.

However, while the execution speed certainly is an interesting evaluation criterion, especially in the context of bin packing heuristics, it cannot be considered the single most important quality aspect. In fact, since the focus of this research lies on finding cost-efficient long-term cloud resource assignments it is not expected that the presented approaches are applied in dynamic reallocation scenario. That is, the heuristics are best suited to assist the decision process when initially setting up a portfolio of cloud instances for a given set of applications. Since no continuous monitoring and optimization of resource usage patterns are required, the execution speeds can be considered to be of lesser relevance. Important is that the problem space is broadly explored.

\begin{figure}[ht]
    \centering
    \begin{subfigure}[b]{\linewidth}
    \includegraphics[width=\linewidth]{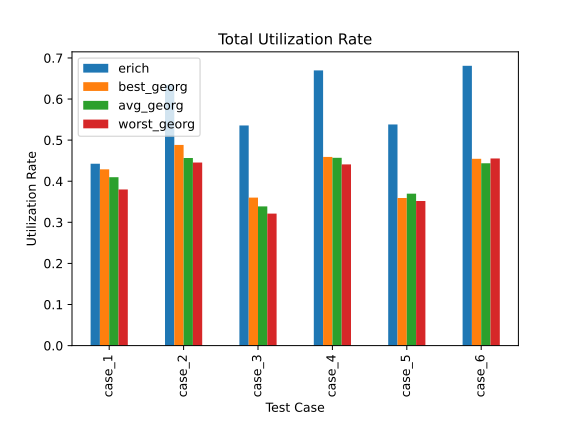}
    \end{subfigure}
\caption{\label{fig:total_utilzation}Instance Utilization Rates}
\end{figure}

Another relevant performance indicator for the proposed optimization heuristics is the packing density. In this work the utilization rate of an instance is defined as the total expected resource demand of all assigned applications over the relevant time slots, relative to the absolute capacity provided for the same time period. The results shown in figure \ref{fig:total_utilzation} clearly show the dominance of \emph{ERICH} over \emph{GEORG} in terms of the average packing density of the allocation instances. It is important to note that the definition of the utilization rate focuses solely on the expected resource demands of the individual applications. The resource demand deviations relevant to the proposed stochastic component of the model are not included in this metric. Therefore, the yield packing density can be considered adequately high.

\begin{figure}[ht]
    \centering
    \begin{subfigure}[b]{\linewidth}
    \includegraphics[width=\linewidth]{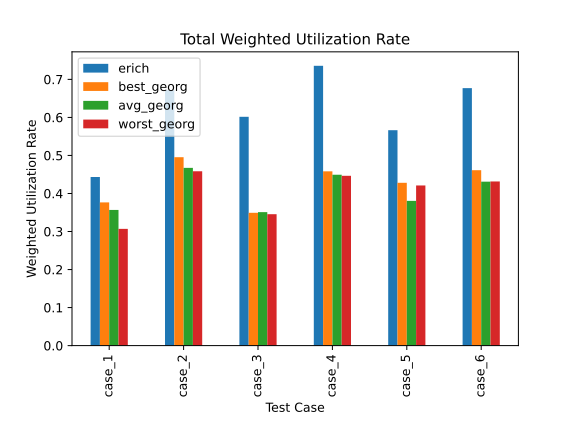}
    \end{subfigure}
\caption{\label{fig:weighted_utilization}Weighted Instance Utilization Rates}
\end{figure}

Figure \ref{fig:weighted_utilization} depicts the data for the weighted utilization rate. Instead of simply taking the average utilization rate over all instances within a portfolio, this metric puts an emphasis on those instances with longer allocation periods. The higher the number of time slots for which an instance is assigned to the portfolio, the more it contributes to the overall utilization rate. However, the conclusion that the greedy optimization approach yields tighter-packed portfolios remains unchallenged.

% \begin{figure}[ht]
%     \centering
%     \begin{subfigure}[b]{.49\linewidth}
%     \includesvg[width=\linewidth]{images/total_utilization.svg}
%     \caption{Total utilization rate}
%     \label{fig:total_utilzation}
%     \end{subfigure}
%     \begin{subfigure}[b]{.49\linewidth}
%     \includesvg[width=\linewidth]{images/weighted_utilization.svg}
%     \caption{Weighted utilization rate}
%     \label{fig:weighted_utilization}
%     \end{subfigure}

% \caption{\label{fig:utilization_rates}Instance utilization rates}
% \end{figure}

\begin{figure}[ht]
    \centering
    \begin{subfigure}[b]{\linewidth}
    \includegraphics[width=\linewidth]{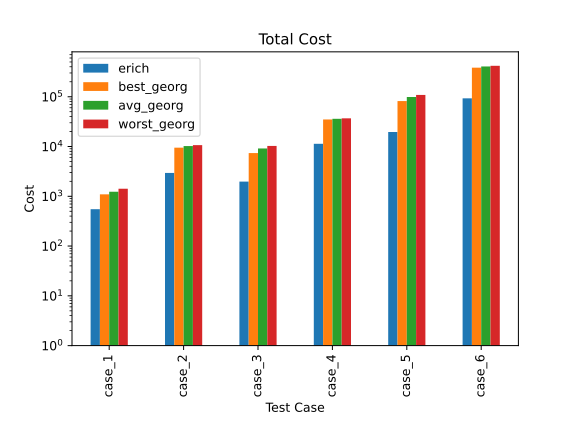}
    \end{subfigure}
\caption{\label{fig:cost_overall}Overall Portfolio Cost}
\end{figure}

Figure \ref{fig:cost_overall} depicts the overall performance of the algorithms in terms of the cost incurred by the resulting packing patterns. Note that the graph shows the results on a logarithmic scale. What can be seen is that \emph{ERICH}, similar to the analysis of the execution speeds and packing density, outperforms the genetic algorithm substantially in each and every of the evaluated test cases. However, the variance within the population of the genetic algorithm is quite high, indicating that the problem space is actually searched rather broadly. Nonetheless, the much simpler approach of the greedy bin packing algorithm shows to be better suited for the evaluated test cases. This can likely be explained by the fact that it acts more target driven in its search for cost-efficient instance types. For the GA, in order to preserve the genetic diversity, a less direct approach was chosen. The crossover operator does not per se assign new instances. Rather, it serves as a smart selection mechanism, merging well-fitting allocations from individuals. New instances are allocated by the insertion heuristic, which is applied to repair broken chromosome. This mechanism, however, relies on a simple semi-random first fit decreasing logic and is not geared towards selecting cost-efficient instances.

% Similar to the analysis of the execution speeds and packing density, the evaluation of the overall cloud portfolio cost shows that the naive \emph{ERICH} outperforms the genetic algorithm by a significant margin, as seen in figure \ref{fig:cost_overall}. However, that is not to say that the GA did not work as intended. Figure \ref{fig:cost_generation} depicts the fitness level over multiple generations for the data set $case\_6$. Starting from an initial population (generation 0) with a very high degree of genetic diversity, the average fitness score continuously improves as the generations progress. In the end, the average cost has been cut by more than half by the time generation 10 has been processed. Moreover, the initial population can also serve as a general benchmark for the quality of the yielded solutions of both algorithms. As mentioned, the GA's population initialization has a strong component of randomness to it. Since both algorithms produce results with notably lower costs compared to the initial population, it can be said that either approach significantly improves upon the solutions one would yield from an unplanned allocation of resources.

\begin{figure}[ht]
    \centering
    \begin{subfigure}[b]{\linewidth}
    \includegraphics[width=\linewidth]{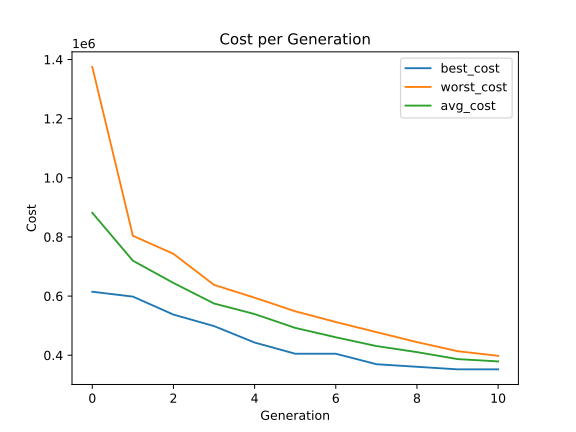}
    \end{subfigure}
\caption{\label{fig:cost_generation}Cost per Generation for Data Set $case\_6$}
\end{figure}

However, that is not to say that the GA did not work as intended. Figure \ref{fig:cost_generation} depicts the fitness level over multiple generations for the data set $case\_6$. Starting from an initial population (generation 0) with a very high degree of genetic diversity, the average fitness score continuously improves as the generations progress. In the end, the average cost has been cut by more than half by the time generation 10 has been processed. As has been shown in section \ref{state_of_the_art} related work in the field of cloud cost optimization often operates under drastically different assumptions. Therefore, the proposed algorithms are not compared to already existing approaches. Instead, the initial population of the genetic algorithm serves as a baseline comparison for the quality of the yielded solutions of both heuristics. As mentioned, the GA's population uses a semi-random first fit mechanism for finding allocations. Since both algorithms produce results with notably lower costs compared to the initial population, it can be said that either approach significantly improves upon the solutions one would yield from such simple and unplanned allocations of resources.

% \begin{figure}[ht]
% \centering
% \begin{minipage}{.5\linewidth}
%   \centering
%   \includesvg[width=\linewidth]{images/cost_overall.svg}
%   \captionof{figure}{Overall portfolio cost}
%   \label{fig:cost_overall}
% \end{minipage}%
% \begin{minipage}{.5\linewidth}
%   \centering
%   \includesvg[width=\linewidth]{images/cost_generation.svg}
%   \captionof{figure}{Cost per generation for data set $case\_6$}
%   \label{fig:cost_generation}
% \end{minipage}
% \end{figure} 

%% file: text/6_conclusion.tex
\section{Conclusion}
\label{conclusion}
In a world where cloud deployment strategies become ever increasingly popular, finding cost-efficient resource allocations is a highly relevant industry problem. In this paper we have presented a formal model for the domain of cloud portfolio management, following an application-centered long-term optimization approach considering uncertain resource demands and incorporating heterogeneous marketplaces. The specification falls into the category of bin packing problems and is characterized by its temporal and stochastic nature. To the best of our knowledge current literature does not provide comprehensive solutions to such long-term probabilistic resource allocation problems. With our research we try to close this gap, while also providing optimization approaches employable by those who wish to find initial cost-efficient cloud portfolio allocations over a longer time horizon.

Two distinct optimization heuristics have been developed for this NP-hard problem, one following a traditional first fit approach, while the other is built on top of the framework of genetic algorithms. The GA uses genetic operators specifically adapted to the problem at hand, using a temporal biased crossover and a mutation procedure that incorporates a temporal dominance criterion as defined by us. However, although the genetic algorithm does explore the broad problem space quite thoroughly, the evaluation has shown that it is significantly outperformed in terms of execution, packing density and overall cost-efficiency.

In future research we plan to extend the problem specification and tweak the optimization heuristics. For instance, we consider it relevant to also look at scenarios where the resource demand of the applications are correlated. Furthermore, in future iterations we also intend to incorporate resource demand load profiles for applications that may change over time. That is, in the current model it is assumed that an application's resource demand is characterized by a normal distribution. More research is necessary to determine the affects on optimization heuristics in case these capacity requirements change over time. 